\documentclass[mathleft
]{an}
\usepackage{graphicx}
\usepackage{amsmath}
\usepackage{amssymb}
\usepackage{times}
\newcommand{\bfb}{\mbox{\boldmath$b$}}
\newcommand{\bff}{\mbox{\boldmath$f$}}
\newcommand{\bfh}{\mbox{\boldmath$h$}}

\newcommand{\bfu}{\mbox{\boldmath$u$}}

\newcommand{\bfB}{\mbox{\boldmath$B$}}

\newcommand{\bfG}{\mbox{\boldmath$G$}}

\newcommand{\bfU}{\mbox{\boldmath$U$}}
\newcommand{\bfX}{\mbox{\boldmath$X$}}

\newcommand\calE{{\cal E}}
\newcommand\calL{{\cal L}}
\newcommand{\bfcalE}{\boldsymbol{{\cal E}}}

\newcommand{\ea}{\textit{et al.\ }}

\begin{document}

\Pagespan{000}{}
\Yearpublication{2010}%
\Yearsubmission{2010}%
\Month{1}%
\Volume{000}%
\Issue{00}%
\DOI{\ \ \ \bf draft version of \today
}%

\title{A Self-Consistent Treatment of the Electromotive Force \\ in Magnetohydrodynamics for Large Diffusivities}

\author{A. Courvoisier\inst{1}
\and D.W. Hughes\inst{1}\fnmsep\thanks{Corresponding author:
  \email{d.w.hughes@leeds.ac.uk}\newline}
\and M.R.E. Proctor\inst{2}
}
\titlerunning{Self-consistent treatment of the electromotive force}
\authorrunning{A. Courvoisier, D.W. Hughes \& M.R.E. Proctor}
\institute{
Department of Applied Mathematics, University of Leeds, Leeds LS2 9JT, UK
\and 
DAMTP, Centre for Mathematical Sciences, University of Cambridge, Cambridge CB3 0WA, UK}



\abstract{
The coupled equations that describe the effect of large-scale magnetic and velocity fields on forced high-diffusivity magnetohydrodynamic flows are investigated through an extension of mean field electrodynamics. Our results generalise those of R\"adler \& Brandenburg (2010), who consider a similar situation but assume that the effect of the Lorentz force on the momentum equation can be neglected. New mean coupling terms are shown to appear, which can lead to large-scale growth of magnetic and velocity fields even when the usual $\alpha$-effects are absent. }

\maketitle

\section{Introduction}

Magnetohydrodynamic turbulence, particularly in the context of astrophysics, is often studied within the framework of mean field electrodynamics, an elegant closure theory that describes the evolution of mean quantities in terms of transport coefficients determined from averaged small-scale properties of the flow and magnetic field (see, for example, Moffatt 1978, Krause \& R\"adler 1980).

The simplest form of the theory considers the case when the magnetic field $\bfB$ can be regarded as kinematic, evolving under the influence of a velocity $\bfU$, but exerting no influence back on the velocity. Only the induction equation is then of significance. The theory proceeds by decomposing the velocity and the magnetic field into mean and fluctuating parts,
\begin{equation}
\bfU = \bfU_0 + \bfu , \qquad
\bfB = \bfB_0 + \bfb ,
\label{eq:meanfluc}
\end{equation}
under an averaging procedure obeying the Reynolds rules. The most common practice is to adopt spatial averaging, assuming a distinct scale separation between that of the large-scale (mean) fields and the small-scale (fluctuating) fields. If, for simplicity, we assume that there is no mean flow (i.e.\ $\bfU_0 =0$), then averaging the induction equation leads to the following equations for the mean and fluctuating magnetic fields,
\begin{equation}
\frac{\partial \bfB_0}{\partial t} = \nabla \times \bfcalE + \eta \nabla^2 \bfB_0 ,
\label{eq:mean_induction}
\end{equation}
\begin{equation}
\frac{\partial \bfb}{\partial t} = \nabla \times \left( \bfu \times \bfB_0 \right)
+ \nabla \times \bfG
+ \eta \nabla^2 \bfb ,
\label{eq:fluc_induction}
\end{equation}
where $\bfcalE = \langle \bfu \times \bfb \rangle$ is the mean electromotive force (emf), $\bfG = ( \bfu \times \bfb ) - \langle \bfu \times \bfb \rangle $ and $\eta$ is the magnetic diffusivity.

Equation (\ref{eq:fluc_induction}) can be expressed as
\begin{equation}
\calL ( \bfb ) = \nabla \times \left( \bfu \times \bfB_0 \right) ,
\label{eq:fluc_op}
\end{equation}
where $\calL$ is a linear operator. If the small-scale field can grow in the absence of a mean field, i.e.\ if the equation $\calL ( \bfb ) = 0$ has exponentially growing solutions, then a small-scale dynamo is operative, and the interpretation of any mean field equations becomes problematic (Cattaneo \& Hughes 2009). If, however, non-decaying solutions for $\bfb$ depend crucially on a non-zero mean field $\bfB_0$, then $\bfb$, and hence $\bfcalE$, depend linearly on $\bfB$, and it is customary to express $\bfcalE$ as a series in $\bfB$ and its spatial derivatives (though see Hughes \& Proctor (2010) for a discussion of the significance of the omitted temporal derivatives), usually written as
\begin{equation}
{\calE}_i = \alpha_{ij}B_{0j} + \beta_{ijk}\frac{\partial B_{0j}}{\partial x_k} + \cdots ,
\label{eq:mean_emf}
\end{equation}
where it is anticipated that the large spatial scale of $\bfB_0$ will lead to rapid convergence. The tensors $\alpha_{ij}$ and $\beta_{ijk}$ depend on the statistical properties of the velocity field and on $\eta$.

The above kinematic formulation  can be regarded as the study of the evolution of weak, long wavelength magnetic field perturbations to a pre-existing non-magnetic flow. One   may then extend this idea so as to consider long wavelength perturbations, in both the magnetic field \textit{and} the velocity, of a \textit{magnetohydrodynamic} [MHD] state involving both small-scale field and flow; here the perturbations are again infinitesimal, but the background state is, in general, fully nonlinear. This problem has been tackled recently in a variety of ways. Courvoisier, Hughes \& Proctor (2010a) considered the evolution of linear, three-dimensional perturbations to two-dimensional MHD basic states resulting from prescribed forcings in the momentum equation in the presence of a background magnetic field with non-zero flux. Courvoisier \ea (2010a) stressed the importance of treating the magnetic and velocity fields on an equal footing, and showed how the linear evolution of the mean field and mean flow are both dependent on terms proportional to both the mean field and the mean flow. In an Appendix, Courvoisier \ea (2010a) outlined the formal extension of the theory to describe perturbations of a three-dimensional, MHD background state. This work is extended in Courvoisier \ea (2010b), which considers in more detail the evolution of long wavelength perturbations to a fully nonlinear three-dimensional, small-scale turbulent state, as may result from the nonlinear saturation of a small-scale dynamo. R\"adler \& Brandenburg (2010) have also recently considered the nature of the emf in MHD turbulence. They made the simplifying assumption that both the basic state and perturbations to it were kinematic, omitting the Lorentz force throughout. However, they allowed a basic MHD state to arise by prescribing an external electromotive force in the induction eq\-uation, in addition to an external body force in the momentum equation. They then considered the influence of a weak mean velocity (but no mean magnetic field), their main result being that the ensuing mean emf has a component proportional to the mean flow. This result had also been noted by Courvoisier \ea (2010a: equations (A.17) and (A.18)).

It is interesting to note that in the model studied by R\"adler \& Brandenburg (2010), it is not actually necessary to make the simplifying assumption of neglecting the Lorentz force. In this paper we therefore extend their analysis by retaining the Lorentz force throughout, thus allowing us to study the evolution of more general MHD states. However, we ignore rotation and any effects of mean field gradients, in order to focus on the most basic effects: the mean emfs and Reynolds and Maxwell stresses that arise owing to imposed mean velocity and magnetic fields. We adopt a slightly different approach to that of R\"adler \& Brandenburg (2010), making the assumption that the magnetic Reynolds number $Rm$ is small, so as to make analytic progress. In order that we can consider a basic MHD state for small $Rm$, we retain the external emf in the induction equation, since, without this, any small-scale field would simply decay. We show that when considering a background MHD state (as opposed to a purely hydrodynamic state), the magnetic and velocity perturbations must, for consistency, be treated on the same footing. Thus, one must consider not just the mean emf in the averaged induction equation, but also the mean stress tensors in the averaged momentum equation.

\section{The mean emf and the mean stress tensor}

Our treatment starts with a basic MHD state consisting of fluctuating, small-scale velocity and magnetic fields, $\bfU$ and $\bfB$, that are driven by an applied body force and an applied electromagnetic forcing. The fluid has kinematic viscosity $\nu$ and magnetic diffusivity $\eta$.  The basic state is described by the following non-dimensional equations,
\begin{equation}
\begin{split}
R_m& \left( \frac{ \partial \bfU}{\partial t} + 
\bfU \cdot \nabla \bfU \right) \\
& = R_m \left(-\nabla \Pi + \chi \bfB \cdot \nabla \bfB \right) + 
P_m \nabla^2 \bfU + \bff, 
\label{eq:mom_bs}
\end{split}
\end{equation}
\begin{equation}
\begin{split}
R_m & \left( \frac{ \partial \bfB}{\partial t} + \bfU \cdot \nabla \bfB \right) \\
&=R_m \left( \bfB \cdot \nabla \bfU \right) + \nabla^2 \bfB + \bfh,
\label{eq:mag_bs}
\end{split}
\end{equation}
where $\Pi$ represents the total pressure. The velocity is scaled with ${\cal U}={\cal F}{\cal L}^2/\eta$ and the magnetic field with ${\cal B}={\cal H}{\cal L}^2/\eta$, where $\cal F$, $\cal H$ and $\cal L$ are representative scales for the applied body force, the applied electromagnetic force and length. The dimensionless parameters are the magnetic Reynolds number $R_m = {\cal U} L/\eta = {\cal F} {\cal L}^3/\eta^2$; the magnetic Prandtl number, $P_m=\nu/\eta$; and $\chi = ({\cal H}^2 / \rho \mu_0)/ {\cal F}^2$, where $\rho$ is the (constant) density and $\mu_0$ the magnetic permeability. The parameter $\chi$, which may be written as ${\cal M}_A^{-2}$, where ${\cal M}_A$ is the Alfv\'enic Mach number, describes the relative importance of the magnetic field and the velocity in the basic state. The situation investigated by R\"adler \& Brandenburg (2010) is recovered by letting $\chi\rightarrow 0$.

We now introduce applied uniform velocity and magnetic fields, $\bfU_0$ and $\bfB_0$, with the aim of evaluating the essential mean quantities in the momentum and induction equations. We assume that the energies in the mean flow and field are of the same order, or smaller, than those of the basic state. If we express the total velocity as $\bfu + \bfU_0$ and the total magnetic field as $\bfb + \bfB_0$, then the governing equations take the form,
\begin{equation}
\begin{split}
R_m& \left( \frac{ \partial\bfu}{\partial t} + 
\bfu \cdot \nabla \bfu + \bfU_0 \cdot \nabla \bfu \right) \\
& = R_m \left(-\nabla \pi + \chi\bfb\cdot\nabla\bfb + \chi\bfB_0\cdot\nabla\bfb\right) \\ 
&\phantom{=}+P_m\nabla^2\bfu +\bff, 
\label{eq:mom2}
\end{split}
\end{equation}
\begin{equation}
\begin{split}
R_m & \left( \frac{\partial\bfb}{\partial t} + \bfu \cdot \nabla \bfb + \bfU_0 \cdot \nabla \bfb \right) \\
&=R_m \left( \bfb \cdot \nabla \bfu +
\bfB_0 \cdot \nabla \bfu \right)+ \nabla^2 \bfb + \bfh .
\label{eq:mag2}
\end{split}
\end{equation}
We suppose that both $R_m$ and the Reynolds number $R_e \equiv P_m R_m$ are small, with $P_m$ of order unity. We may thus expand $\bfu$ and $\bfb$ in powers of $R_m$, namely 
\begin{equation}
\bfu = \bfu^{(0)} + R_m \bfu^{(1)} + \cdots, \quad
\bfb = \bfb^{(0)} + R_m \bfb^{(1)} + \cdots .
\end{equation}

Similarly, we may expand the fields $\bfu' = \bfu - \bfU$ and $\bfb' = \bfb - \bfB$, induced by the imposed mean velocity and magnetic fields, as
\begin{equation}
\bfu' = R_m \bfu'^{(1)} + \cdots , \quad
\bfb' = R_m \bfb'^{(1)} + \cdots,
\end{equation}
where
\begin{align}
P_m \nabla^2\bfu'^{(1)} &= ({\bfU_0}\cdot\nabla\bfu^{(0)} - \chi {\bfB_0}\cdot\nabla\bfb^{(0)}),
\label{eq:low1}\\
\nabla^2\bfb'^{(1)} &= ({\bfU_0}\cdot\nabla\bfb^{(0)}-{\bfB_0}\cdot\nabla\bfu^{(0)}) .
\label{eq:low2}
\end{align}
The right hand side of (\ref{eq:low1}) is solenoidal, and hence there is no pressure gradient term. We assume that there is no mean emf in the basic state, described by equations~(\ref{eq:mom_bs}) and (\ref{eq:mag_bs}); hence the leading order contribution to the emf is given by 
\begin{equation}
\boldsymbol{\cal E} = R_m \langle \bfu^{(0)}\times\bfb'^{(1)} + \bfu'^{(1)}\times\bfb^{(0)} \rangle .
\label{eq:emf_first_order}
\end{equation}
To aid calculation, we now further suppose that $\bff$ and $\bfh$ are independent of $t$, spatially periodic and  monochromatic, so that
\begin{equation}
\nabla^2 \bff = -\lambda^2 \bff, \qquad \nabla^2 \bfh = -\lambda^2 \bfh,
\label{eq:mono}
\end{equation}
with $|\lambda|\sim {\cal L}^{-1}$. It can be shown that $\bfu^{(0)}$ and $\bfb^{(0)}$ are then also monochromatic. Hence we can write
\begin{align}
\langle \bfu^{(0)} \times \bfb'^{(1)} \rangle & =
-\lambda^{-2} \langle \nabla^2 \bfu^{(0)} \times \bfb'^{(1)} \rangle \label{eq:emf_a} \\ 
& = -\lambda^{-2} \langle \bfu^{(0)} \times \nabla^2 \bfb'^{(1)} \rangle ,
\label{eq:emf_b}
\end{align}
and similarly for the other term in (\ref{eq:emf_first_order}). All surface terms involved in the integrations by parts needed to transform (\ref{eq:emf_a}) to (\ref{eq:emf_b}) vanish, by assumption of spatial periodicity. Then using equations~({\ref{eq:low1}) and (\ref{eq:low2}) we obtain exactly,
\begin{equation}
\begin{split}
\lambda^2 R_m^{-1} {\cal E}_i &= \bfU_0 \cdot \langle (1-P_m^{-1}) \epsilon_{ijk} \nabla u_j^{(0)} b_k^{(0)} \rangle \\& + \bfB_0 \cdot \langle \epsilon_{ijk} ( u_j^{(0)} \nabla u_k^{(0)} + P_m^{-1} \chi \nabla b_j^{(0)} b_k^{(0)}) \rangle .
\label{eq:indterm}
\end{split}
\end{equation}
The terms in $\bfB_0$ may be recognised as the usual kinetic and magnetic $\alpha$-effects, while the term in $\bfU_0$ has the same form as that discussed by R\"adler \& Brandenburg (2010) in a more general context. 

Thus far there is little that is new. However, crucially, we can see that because the momentum equation now contains $\bfB_0$, through the Lorentz force, we also have a non-trivial expression for the mean stress tensor, $R_{ij}$, at leading order, namely
\begin{equation}
\begin{split}
R_m^{-1}R_{ij} = \langle u'^{(1)}_i u^{(0)}_j &+ u^{(0)}_i u'^{(1)}_j \rangle \\ &-
\chi \langle b'^{(1)}_i b^{(0)}_j + b^{(0)}_i b'^{(1)}_j \rangle .
\label{eq:stress}
\end{split}
\end{equation}
Under the assumptions (\ref{eq:mono}), and via manipulations analogous to those leading to (\ref{eq:indterm}), we obtain
\begin{equation}
\lambda^2 R_m^{-1} R_{ij} = \chi \bfB_0 \cdot \langle ( 1+P_m^{-1} )
( \nabla b^{(0)}_i u^{(0)}_j + \nabla b^{(0)}_j u^{(0)}_i ) \rangle .
\label{eq:momterm}
\end{equation}
It might have been expected that there would also be a term proportional to $\bfU_0$ in this expression. Were such a term to exist, it would be analogous to the AKA (anisotropic kinetic $\alpha$) effect introduced by Frisch, She \& Sulem (1987). There are such contributions present here, but they cancel out; in fact it can be shown, in general and not just for monochromatic velocity fields, that the AKA effect vanishes in the first order smoothing approximation, for steady forcing as assumed. However, a non-zero contribution may be expected for more general forcings.  In spite of this, it is plain from the form of the terms in expressions~(\ref{eq:indterm}) and (\ref{eq:momterm}) that the evolution equations for the large-scale velocity and magnetic fields will generically be coupled together. Interestingly, both types of coupling term depend on the tensor
\begin{equation}
Q_{ijk}=\left\langle\frac{\partial u^{(0)}_i}{\partial x_j} b^{(0)}_k\right\rangle=-\left\langle{\frac{\partial b^{(0)}_k}{\partial x_j} u^{(0)}_i}\right\rangle .
\label{eq:tensor}
\end{equation}

The above calculations have been performed for uniform, steady $\bfU_0$ and $\bfB_0$. However, and this is the underlying principle of mean field electrodynamics, the results for the mean emf and the mean stress tensor can be used to determine the evolution of velocity and magnetic fields depending slowly on space and time. Even for small $Rm$, if the length scales are sufficiently long then the terms involving first derivatives dominate over diffusion terms, which may therefore be neglected. In this case, the mean momentum and induction equations take the form
\begin{equation}
\partial_T U_{0i} + \partial_{X_j}R_{ij} = -\partial_{X_{i}} P_0,
\end{equation}
\begin{equation}
\partial_T B_{0i} = \epsilon_{ijk} \partial_{X_{j}} \calE_k ,
\end{equation}
where $T$ and $\bfX$ are long time and space variables.
We now ignore the familiar $\alpha$-effect terms, and look for solutions proportional to $e^{i\boldsymbol{K}\cdot\boldsymbol{X} +sT}$. After appropriate scaling of the time and space variables, to absorb the factors of $\lambda^2 R_m^{-1}$, and elimination of the pressure using $\nabla\cdot\bfU_0=0$, we obtain (writing $W^\pm_{ijk}=Q_{ikj}\pm Q_{kij}$)
\begin{align}
sB_{0i}&=\left(1-P_m^{-1}\right)W^-_{ikl}\cdot iK_k U_{0l} , \\
sU_{0i}&=\left(1+P_m^{-1}\right)\chi S_{ij}W^+_{jkl}\cdot i K_k B_{0l},
\end{align}
where $S_{ij}=\delta_{ij}-K_iK_j/|\boldsymbol{K}|^2$.
While $W^+$, being symmetric in its first two arguments, will vanish in isotropic situations, there is no reason for it to vanish in general. Indeed, for a simplified problem involving a 2D basic state, Courvoisier \ea (2010a) have exhibited cases where this term is non-zero. Because the equations are coupled, exponential growth is possible even without the $\alpha$-effect.  The square of the growth rate $s$ is given as the eigenvalue of the matrix
\begin{equation}
- \chi \left(1-\frac{1}{P_m^2} \right) W^-_{ikl} K_k K_p S_{lm}W^+_{mpq}, \label{eq:meanmatrix}
\end{equation}
and it appears that solutions with a positive real part of $s$ can be found in a wide variety of situations, provided that $P_m\ne 1$.

\section{Discussion}

Starting from a small-scale, MHD basic state, forced via a body force and an externally applied emf, we have considered the mean emf, $\bfcalE$, and the mean Reynolds-Maxwell stress tensor, $R_{ij}$, resulting from the imposition of spatially uniform, steady, velocity and magnetic fields. Under the assumption that the fluids and magnetic Reynolds numbers are small, we have obtained analytic expressions for $\bfcalE$ and $R_{ij}$, (\ref{eq:indterm}) and (\ref{eq:momterm}). These are linear in the mean fields, with transport coefficients dependent on the means of various quadratic fluctuating quantities of the basic state.

The main point of our paper is that in a true MHD state, in which neither the velocity nor the magnetic field can be regarded as dominant, it is imperative that the velocity and magnetic fields are treated on an equal footing. This is in sharp contrast to traditional mean field electrodynamics, which is essentially a kinematic theory, describing the evolution of a (formally weak) large-scale magnetic field under the influence of a prescribed velocity. It is then only the mean emf that is of interest; furthermore, the mean emf depends only on the fluctuating velocity field. In the case considered in this paper, however, $\bfcalE$ and $R_{ij}$ are of equal importance and, moreover, they both depend on both the fluctuating velocity and the fluctuating magnetic field. We believe that this is an important message, in that it highlights the shortcomings in, for example, considering only the emf, and then only the part proportional to the mean magnetic field (the $\alpha$-effect), for MHD turbulence. Although it is often claimed that the $\alpha$-effect in the nonlinear (MHD) regime can be expressed as the difference between the flow helicity and the current helicity (essentially the $\bfB_0$ term in (\ref{eq:indterm})), our analysis shows clearly that this tells only part of the story. In the general case, the coupling between the mean velocity and magnetic fields is such that the growth of a large-scale disturbance can be expected even when the usual kinematic and magnetic $\alpha$-effects vanish. Such a mode of instability  is not accessible if the influence of the Lorentz force on the fluid momentum is dropped.

\acknowledgements
This work was supported by the UK Science and Technology Facilities Council.



\begin{thebibliography}{}
  \bibitem{} Cattaneo, F., Hughes, D.W.: 2009, MNRAS~395, L48
  \bibitem{} Courvoisier, A., Hughes, D.W., Proctor, M.R.E.: 2010a, RSPSA~466, 583
  \bibitem{} Courvoisier, A., Hughes, D.W., Proctor, M.R.E.: 2010b, in preparation
  \bibitem{} Frisch, U., She Z.S., Sulem P.-L.: 1987, Physica D~28, 382
  \bibitem{} Hughes, D.W., Proctor, M.R.E.: 2010, PhRvL~104, 024503
  \bibitem{} Krause, F., R\"adler, K.-H.: 1980, \textit{Mean-Field Magnetohydrodynamics and Dynamo Theory}, Pergamon Press, Oxford
  \bibitem{} Moffatt, H.K.: 1978, \textit{Magnetic Field Generation in Electrically Conducting Fluids}, Cambridge University Press, Cambridge
  \bibitem{} R\"adler, K.-H., Brandenburg, A.: 2010, AN~331, 14

\end{thebibliography}
\end{document}